\documentstyle[emulateapj]{article}
\include{psfig}

\newcommand{\ie}{{\sl i.e.}\ }
\newcommand{\eg}{{\sl e.g.,}\ }
\newcommand{\etal}{{\sl et al.~}}

\begin{document}

\title{HST observations of the LMC field around SN~1987A: distance
determination with Red Clump and Tip of the Red Giant Branch stars}

\author{M. Romaniello\altaffilmark{1}, M. Salaris\altaffilmark{2,4}, 
S. Cassisi\altaffilmark{3,4}, N. Panagia\altaffilmark{5}}

\lefthead{Romaniello \etal}
\righthead{Red Clump and TRGB distance to the LMC} 

\altaffiltext{1}{ESO, Karl-Schwarzschild-Stra{\ss}e 2, D-85748
Garching bei M\"unchen, Germany; \mbox{{\it mromanie@eso.org}}}
\altaffiltext{2}{Astrophysics Research Institute, Liverpool John Moores 
University, Twelve Quays House, Egerton Wharf, Birkenhead CH41 1LD, UK; 
{\it ms@staru1.livjm.ac.uk}} 
\altaffiltext{3}{Osservatorio Astronomico di Collurania, Via M. Maggini,
64100 Teramo, Italy; {\it cassisi@astrte.te.astro.it}} 
\altaffiltext{4}{Max-Planck-Institut f\"ur Astrophysik,
Karl-Schwarzschild-Stra{\ss}e 1, D-85748 Garching bei M\"unchen, Germany}
\altaffiltext{5}{Space Telescope Science Institute, 3700 San Martin
Drive, Baltimore, MD~21218, USA; on assignment from the Astrophysics
Division, Space Science Department of ESA; {\it panagia@stsci.edu}} 

\setcounter{footnote}{0}
\begin{abstract}

We have used HST-WFPC2 multiband observations of a field around
SN~1987A in the Large Magellanic Cloud to measure its distance from
the Sun. The observations allowed us to
carefully determine the interstellar extinction along the line of
sight to a large number of stars and to measure the LMC distance by using
two stellar distance indicators: the Red Clump and the Tip of the Red
Giant Branch. From an application of the Red Clump method we obtain a
distance modulus $(m-M)_{0,RC}^{LMC}$=18.59$\pm$0.04$\pm$0.08~{\sl mag}
(statistical plus systematic error), in good agreement with the
distance derived by using the Tip of the Red Giant Branch stars,
namely $(m-M)_{0,TRGB}^{LMC}$=18.69$\pm$0.25$\pm$0.06~{\sl mag} (statistical
plus systematic error). Both values agree well with the distance to
the SN~1987A as determined from a study of its inner ring fluorescent
echo ($(m-M)_{SN~1987A}$=18.55$\pm$0.05~{\sl mag}, Panagia 1998), thus 
excluding distance moduli lower than 18.43 to a 99.7\% significance 
level. Differences with respect to previous results obtained using the
same distance indicators are discussed. 
\end{abstract} 

\noindent
{\em Subject headings:} galaxies: distances and redshifts --- 
galaxies: individual (Large Magellanic Cloud) ---
stars: evolution

\section{Introduction}

The distance to the Large Magellanic Cloud is a fundamental step in
the cosmological distance ladder: since the Cepheid extragalactic
distance scale is tied to the LMC distance, any error in the
determination of the distance to the LMC propagates directly to the
cosmological distances. 

Recent determinations based on the light echoes of SN~1987A (Panagia
1998), on the HIPPARCOS calibrated RR Lyrae (via Main Sequence fitting
technique) and Cepheids distance scale (Gratton \etal 1997, Reid
1997, Feast \& Catchpole 1997, Oudmaijer \etal 1998), on the
theoretical calibration of RR Lyrae and Tip of the Red Giant Branch
(TRGB) stars brightness (Salaris \& Cassisi 1998) provide distance
moduli ranging approximately between $(m-M)_{0}^{LMC}$=18.50 and
18.70~{\sl mag} (``long'' distance scale). On the other hand, the
straightforward application of the Red Clump (RC) method (Paczynski \& Stanek
1998) for distance determinations to two LMC fields by Stanek,
Zaritsky \& Harris (1998) provides a much shorter distance, namely
$(m-M)_{0}^{LMC}$=18.065$\pm$0.031$\pm$0.09~{\sl mag} (statistical plus
systematic error).
According to Cole (1998) and Girardi \etal (1998), when taking properly into
account population effects on the RC luminosity, that distance modulus has to
be increased by $\sim$0.2-0.3~{\sl mag}, making it only marginally consistent
with the ``long'' distance scale. It is unpleasant to notice how different
independent stellar distance indicators provide different answers
about such an important quantity. 

In this paper we make use of multicolor HST-WFPC2 observations
(Romaniello 1998, Romaniello \etal 1999) of a circular region with a
radius of approximately 2\arcmin\ centered on the SN~1987A, obtained as
part of the long term General Observer {\bf S}upernova {\bf IN}tensive
{\bf S}tudy (SINS) project. The main aim of our investigation is to
provide an accurate distance determination to this field by using
two independent stellar standard candles mentioned above: the TRGB and
the RC. Since the observed stellar field is located around the
SN~1987A we can perform a meaningful comparison between the derived
distance modulus and the distance to the supernova as independently
determined by means of studies of the fluorescence echoes from the Supernova
circumstellar ring (Panagia 1998). Moreover, the availability of
multiband observations has allowed a careful and homogeneous reddening
determination by means of a newly developed technique (Romaniello
1998, Romaniello \etal 1999). This constitutes an important
improvement over previous works, because it permits to carefully take
into account the existing small scale variations in the internal
extinction of the observed LMC field (which is located in an area
containing a large number of early type stars interspersed with HII
regions and Supernova remnant shells). 

In \S 2 we briefly discuss the observational data and the technique
employed for the reddening determination. Section 3 deals with the
distance determinations, while in \S 4 we discuss the main results. 

\section{The data}

Since 1994, Supernova~1987A has been imaged every year with the Wide
Field and Planetary Camera~2 (WFPC2) on board the NASA/ESA Hubble
Space Telescope (HST) in the context of the long-term SINS project
(PI: Robert P. Kirshner). 
Here, we use the observations taken on September~24, 1994,
February~6, 1996 and July~10 1997. They provide full coverage of
a circular portion of the LMC with a radius of roughly 2\arcmin\
($\sim30$~pc) centered on SN~1987A in six wide band filters covering the
spectral region from the ultraviolet to the near infrared: F255W,
F336W, F439W, F555W, F675W and F814W. 

The full description of the data and of the reduction process is given
elsewhere (Panagia \etal~1999, Romaniello \etal~1999). In brief, the
observations were processed through the standard PODPS (Post
Observation Data Processing System) pipeline for bias removal and flat
fielding. In all cases the available images for each filter were
combined to remove cosmic rays events. 
The fluxes were measured performing aperture photometry following the
prescriptions by Gilmozzi (1990) as refined by Romaniello
(1998), \ie measuring the flux in a circular aperture of 2~pixels
radius and the sky background value in an annulus of inner radius
3~pixels and width 2~pixels. The flux calibration is obtained using
the internal calibration of the WFPC2 (Whitmore~1995) which is
typically accurate to within $\pm$ 5\%. We use the spectrum of Vega as
photometric zero point.

Both the distance determination methods used in this paper rely solely
on the luminosity of the Red Giant stars in the F814W filter. The
WFPC2 sensitivity is extremely stable in this spectral region. The
zero point variation, as evaluated from the PHOTFLAM keyword in the
image headers, is smaller than 3\% over the entire time-span covered
by our observations. 
The uncertainty on the final calibrated zero points of the photometry is of
the order of $\pm$0.04 mag.

In our analysis we have taken full advantage of the wealth of information
provided by the broad wavelength range (about 2300-9000 \AA) covered by the
observations. By means of multiband fits with the theoretical spectra by
Bessel, Castelli \& Plez (1998), we have derived both the intrinsic stellar
properties and the interstellar extinction along the line of sight to the
individual stars (Romaniello 1998 and Romaniello \etal 1999). In our field,
we find that the hot ($T_{eff}\gtrsim 10,000$~K, \ie young) and cold
($6,500\lesssim T_{eff}\lesssim 8,500$~K, \ie presumably old) Main Sequence
stars are affected, on average, by the same amount of extinction. When no
determination of the reddening for a given star on the Red Giant branch was
possible, the mean value from its neighbors belonging to the same
intermediate-old stellar population was used. As recently noticed also by
Zaritsky (1999), one has to be extremely careful that the stars from which the
extinction is determined belong to the same population as the stars one is
studying to take into account possible population-dependent effects.

The resulting dereddened F814W$_0$~vs.~(F555W$-$F814W)$_0$ Color-Magnitude
Diagram (CMD) is shown in
panel~(a) of Figure~1. In panel~(b) of the same Figure we show the
reddening distribution\footnote{The extinction in the various HST filters has
been translated into E(B$-$V) according to the reddening law as determined by
Scuderi \etal~(1996).} {\sl individually measured for 2510 stars} in our field
(one every 13 square arcsec, on average). The peak occurs at
E(B$-$V)$=0.20$~{\sl mag}, and the distribution displays non-negligible
scatter: $\sigma(\mathrm{E(B-V)})=0.072$~{\sl mag} rms, at least twice the
measurement errors. It is clear that in a
case like this an improper evaluation of the interstellar extinction may
introduce an error/uncertainty in the subsequent distance modulus by
as much as $\pm0.14$~{\sl mag} ($rms$). By measuring it directly for
individual stars or, in the worst case, from numerous well measured
neighbors, we eliminate this uncertainty that can significantly affect
the final result. 


\section{Distance determination}

An inspection to the CMD displayed in Figure 1 (panel a) reveals the
presence of an intermediate-old stellar population in the red part of
the diagram. The RC ($17\lesssim$F814W$_0\lesssim18$~{\sl mag},
$1\gtrsim$(F555W$-$F814W)$_0\gtrsim0.8$~{\sl mag}) and an extended RGB
(F814W$_0\lesssim20$~{\sl mag}, (F555W$-$F814W)$_0\gtrsim0.7$~{\sl mag})
are clearly visible. Typically, our photometry for RGB stars is accurate to
better than 1\% in both filters.

\subsection{The Red Clump} 

The RC is a common feature in many CMDs: it is populated by low-mass,
metal rich stars experiencing central He-burning, and represents the
intermediate-age, metal-rich counterpart of the globular clusters
Horizontal Branch. The I-(V-I) CMD from HIPPARCOS data shows clearly
the local Red Clump, extending horizontally in the $(V-I)_{0}$ interval
approximately between 0.8 and 1.25~{\sl mag}, with a mean absolute magnitude
$M_{I}^{0}=-0.23\pm0.03$~{\sl mag} and a dispersion
$\sigma_{RC}$=0.20~{\sl mag} (Stanek \& Garnavich 1998). 

The constancy of the RC mean brightness over such a wide color range
was interpreted as evidence that it can be used as a stellar standard
candle, independent of the properties of the underlying stellar
population, at least for $(V-I)_{0}$ between 0.8 and
1.25~{\sl mag} (see, \eg Paczynski \& Stanek 1998). However, using
evolutionary stellar models, Cole (1998) and Girardi \etal (1998)
have shown that $M_{I}^{0}$ of the RC does depend on the properties of
the stellar population. In particular, Girardi \etal (1998) have
demonstrated that theoretical stellar models are able to reproduce the
structure and the constancy of $M_{I}^{0}$ with color for the local
HIPPARCOS RC, and that $M_{I}^{0}$ is not a constant among different
populations, but depends on their metallicities. On the observational side,
Twarog, Twarog \& Bricker (1999) have found a dependence of $M_{I}^{0}$ on the
metallicity from the determination of the distance to 2 Galactic open clusters
with ages and metallicities typical of the LMC stellar population; they used
the Main Sequence fitting technique to estimate the distances, by employing
theoretical isochrones calibrated on HIPPARCOS subdwarfs. 
 
In Figure 2 (left panel) we show the RC region in our CMD. 
The dereddened data in the F555W and F814W band have been transformed into the
VI Johnson-Cousins system following Holtzman~\etal (1995). These
transformations are consistent with those derived by convolving the
Bessel~\etal (1998) synthetic spectra with the HST and Johnson-Cousins filters
using the IRAF-STSDAS {\sl synphot} package. These corrections are typically
of 0.03~{\sl mag}, and in all cases smaller than 0.05~{\sl mag}.

We have applied the RC method as described, for
example, in Stanek \etal (1998) by selecting the stars in the range
0.8$<(V-I)_{0}<$1.25~{\sl mag} (note that the RC is almost completely
contained within this color-range. In addition, we have also verified
that the final result does not change even if we include the bluemost
part with $(V-I)_{0}<0.8$~{\sl mag}) and
17.0$<I_{0}<$19.0~{\sl mag}, and fitting the distribution of stars
as a function of the I-band magnitude with the following function (Stanek
\& Garnavich 1998):

$$ n(I_{\rm 0}) = a + b(I_{0}-I_{0,m}) + c(I_{0}-I_{0,m})^{2} + $$
\begin{equation}
          \frac{N_{RC}}{\sigma_{RC}\sqrt{2\pi}}\exp 
          \left(-\frac{(I_{0}-I_{0,m})^{2}}{2\sigma_{RC}^{2}}\right)
\end{equation}

The first three terms correspond to a fit to the 
distribution of RGB stars, while the Gaussian term
represents a fit to the RC. We find the peak magnitude of the RC
population to be $I_{0,m}$=18.12$\pm$0.02~{\sl mag}, while the dispersion
turns out to be $\sigma_{RC}$=0.16~{\sl mag}. The result of the fit is shown
in Figure 2 (right panel). By using $M_{I}^{0}=-0.23\pm0.03$~{\sl mag} for
the local clump and without any evolutionary correction, one would get
$(m-M)_{0,RC}^{LMC}$=18.35$\pm$0.04~{\sl mag} (statistical error only). 


It is important to consider at this point the correction ($\Delta M_{I}$)
due to population effects. The red boundary of the RC in our
CMD is located at $(V-I)_{0}\approx$1.0~{\sl mag}, approximately
0.2~{\sl mag}
bluer than the local HIPPARCOS RC, and the color extension is of
about 0.3~{\sl mag} in $(V-I)$, \ie about 0.1~{\sl mag} less extended than the
local RC.
Using the models from Girardi \etal (1998), the position and color extension
of the RC in the observed LMC field indicates a metallicity ranging between
Z$\approx$0.002 and Z$\approx$0.008 ([M/H]$\approx-1.0 \div -0.4$). A
similar result is derived by using different
theoretical models, such as the ones by Cassisi, Castellani \& Straniero
(1994) or Seidel, Demarque \& Weinberg (1987). 

The value of $\Delta M_{I}$ to be applied to $M_{I}^{0}$ as derived by
Twarog \etal (1999) for a metallicity of [Fe/H]=-0.8 amounts
to -0.31~{\sl mag}. This figure is obtained considering stars in a cluster,
which means stars belonging to a stellar population with a single
metallicity and single age. Cole (1998) and Girardi \etal (1998)
considered a composite stellar population (as the one observed in the
LMC fields) with realistic assumptions about the star formation
history (SFH). Cole (1998) obtained  $\Delta M_{I}=-0.32$ by considering a
SFH as in Holtzman \etal (1997), and
$\Delta M_{I}=-0.23$ when assuming the more ``burst-like'' SFH from
Vallenari \etal (1996). Girardi \etal (1998) have derived $\Delta
M_{I}=-0.23$ for a constant Star Formation Rate in the last 3 Gyr and
equiprobable metallicities between Z=0.004 and Z=0.008, while $\Delta
M_{I}=-0.17$ for the Vallenari \etal (1996) SFH. 

Based on these results we will use an average value $\Delta
M_{I}$=-0.24, adding to the error budget on the final distance modulus
a systematic error of $\pm$0.08~{\sl mag} which takes into account the
uncertainties on $\Delta M_{I}$ coming from the adopted stellar models,
the assumed SFH and the error on the zero point of the photometry. 

The final value for the distance to the observed LMC field is 

$$(m-M)_{0,RC}^{LMC}~=~18.59\pm0.04\pm0.08~mag$$
\centerline {\sl (statistical ~plus ~systematic ~error)}

\subsection{The Tip of the Red Giant Branch}

The use of the TRGB as a distance indicator is discussed at length in
Lee \etal (1993), Madore \& Freedmann (1995), and Salaris \& Cassisi
(1997, 1998). Stars at the TRGB are experiencing the core
Helium-flash, and their luminosity is remarkably constant for a large
range of masses (corresponding to ages equal or larger than $\sim2$
Gyr). Moreover, the absolute I magnitude of TRGB stars is
very weakly affected by the metallicity of the underlying stellar
population, at least for metallicities lower than half-solar (Salaris \&
Cassisi 1997, 1998). The
basic idea of the TRGB method for distance determination is to derive
the position of the TRGB from the observed Luminosity Function (LF) of
the upper RGB population, and to compare it with prescriptions from
theoretical stellar models. As discussed in Salaris \& Cassisi (1998),
all theoretical models agree quite well with each other on the
predicted luminosity of the TRGB. Also the uncertainties on the
theoretical bolometric corrections appear to be quite small. 

The position of the observed TRGB has been determined according to the
procedure described in Lee \etal (1993) and Madore \& Freedman
(1995). We have computed the  differential LF for $I_{0}\leq$17.5, so as
to avoid substantial contamination of RC stars. Due to the
limited spatial extension of the observed field the upper part of the
RGB cannot be very populated ($\sim150$ stars in the selected
brightness range). We have employed bins $\pm$0.25~{\sl mag} wide in the LF,
basing our choice on the results from Monte-Carlo simulations
performed using the theoretical models by Salaris \& Cassisi (1998).
According to these simulations, this bin selection ensures to have always a
RGB population more than two sigma different from zero in the bin
centered on the TRGB brightness. The kernel [-1, 0, +1] (Madore \&
Freedman 1995) has been convolved with the observational LF (our results do
not change appreciably when using a kernel covering a wider baseline,
namely [-1, -2, 0, +2, +1]); the kernel response reflects the gradient
detected across a three-point interval and produces a maximum at the
luminosity where the count discontinuity is the largest. We used the
midpoint of the corresponding luminosity bin as the value of the TRGB
brightness (see top panel of Figure 3). The TRGB in the observed CMD
is located at $I_{0}^{TRGB}$=14.50$\pm$0.25~{\sl mag}. 
From the previous discussion, it is clear that an error bar by
$\pm$0.25~{\sl mag} corresponds to an estimate of the maximum error on the
TRGB position. This value of $I_{0}^{TRGB}$ is in good agreement with the
value of $I_{0}^{TRGB}$=14.53$\pm$0.05~{\sl mag} derived by Reid \etal (1987)
from observations of a large area of the Shapley Constellation III
within the LMC. This value was used in subsequent analyses
(Lee \etal 1993, Salaris \& Cassisi 1997, 1998) for deriving the TRGB
distance to LMC. 


By considering a mean metallicity [M/H]=-0.7 as for the RC stars and
the theoretical TRGB absolute I magnitude from Equation 5 in Salaris
\& Cassisi (1998), one derives 

$$(m-M)_{0,TRGB}^{LMC}~=18.69\pm0.25 ~mag$$ 
\centerline {\sl (statistical ~error ~only)}\\ 

Note that the error introduced by an uncertainty (or a spread) in metallicity
even as large as a factor of 2 is negligible with respect to the error on the
TRGB position. 

In Figure 3 (bottom panel) we show the comparison between the
observational and theoretical LF for the adopted mean value of the
distance modulus and metallicity. The faintest, and most populated, bin in the
observational LF has been used to normalize the population of the
theoretical one. It is comforting  to see how well the theoretical LF
reproduces the observed one over the last 3 magnitudes below the TRGB.
After performing a least square fit we found that the slopes of the
two LF agree within the statistical error. This result confirms also
the negligible level of contamination from different stellar
populations, both in the Galaxy and in the LMC itself (Asymptotic Giant
Branch stars). In Figure 3 the LF for the ``short''
distance scale is also included, namely for
$(m-M)_{0,TRGB}^{LMC}$=18.10~{\sl mag}; it is clear that such a short
distance is ruled out by our data not only because it
predicts RGB stars at magnitudes brighter than the TRGB discontinuity,
but also because it is clearly inconsistent with the remaining part of
the observed LF. 

By adding to the final value of the TRGB distance modulus a
systematic uncertainty of $\pm$0.05~{\sl mag} due to theoretical
uncertainties on the calibration of the absolute TRGB luminosity and
bolometric corrections (Salaris \& Cassisi 1998), and the error
on the zero point of the photometry, we obtain 

$$(m-M)_{0,TRGB}^{LMC}~=~18.69\pm0.25\pm0.06 ~mag$$ 
\centerline {\sl (statistical ~plus ~systematic ~error)}


\section{Discussion}

The distances we obtained from the RC method and the TRGB for the
stellar population around SN~1987A are in good mutual agreement.
When combined, they rule out distance moduli smaller than 18.34 at a
3~sigma level.
They are also in good agreement with the distance to the SN~1987A as
determined by Panagia (1998), namely
$(m-M)_{SN~1987A}$=18.55$\pm$0.05~{\sl mag}. 
Let us note that this value is also
consistent with the one derived from the fit of
theoretical models to the observed Zero Age Main Sequence 
(Romaniello 1998, Romaniello \etal 1999). In conclusion,
these results all agree on a value 
around 18.57 and exclude values lower than 18.43~{\sl mag} to 99.7\%, \ie
3~sigma, confidence level.

Our derived  value of $I_{0}^{TRGB}$ compares well with the results by
Reid \etal (1987) from observations of a different, more extended LMC
field. As a consequence, the distance modulus
derived by Salaris \& Cassisi (1998), who used the Reid \etal (1987)
data together with the same theoretical calibration we employed, agrees
well with our results. Moreover, we have found that the LF of the
upper RGB agrees quite well with theoretical models and, by itself,
rules out distances as short as $(m-M)_{0,TRGB}^{LMC}$=18.10~{\sl mag}. 

The distance modulus we get from the RC method is about 0.5~{\sl mag} higher 
than the value determined by Stanek \etal (1998). About half of this
discrepancy is due to the correction $\Delta M_{I}$ for population
effects we have applied, while the other half is due to an intrinsic
difference in the observed $I_{0,m}$ values. The RC position in our
CMD differs substantially from the results by Stanek \etal (1998);
more precisely, we derive a value for $I_{0,m}$ dimmer by $\approx$
0.3~{\sl mag} and a $(V-I)_{0}$ color redder by $\simeq$ 0.15~{\sl mag}.

An obvious possibility to explain the apparent discrepancy both in magnitude
and color is an improper reddening correction.
We are confident about our treatment of the extinction because it is based on 
individual determinations for a large number of stars in the sample, mostly belonging to
the old population (one
every 13 square arcsecond; \eg Romaniello 1998). As we have already pointed out in Section~2, the
reddening corrections we have applied are the appropriate ones for the old population to which the
Red Giant stars belong. Moreover, in our field the
mean reddening is in good agreement with the independent determination of
the reddening towards SN~1987A as discussed in Scuderi \etal (1996)
which is based on the detailed study of the HST-FOS UV and optical spectrum of
``Star 2", one of the two companion stars near SN~1987A. 
In order to eliminate the discrepancy in the observed $I_{0,m}$ values
one should therefore invoke a $\delta$ E(B$-$V) $\simeq 0.15$ systematic overestimate
of the reddening by Stanek \etal (1998). This seems to be indeed the case,
according to the recent analysis by Zaritsky (1999). He 
finds the existence of population-dependent extinction properties in the LMC,
and concludes that the extinction map derived by Harris, Zaritsky \& Thompson (1997)
and used by
Stanek \etal (1998) is not an accurate representation of the reddening 
to RC stars. Moreover, the real extinction for the RC population
in the regions selected by Stanek \etal (1998) results to be $A_{I}\simeq 0.06$,
which increases the observed $I_{0,m}$ by $\simeq 0.25$~{\sl mag}.
With this correction the level of the RC (as well as its colour) 
in the fields considered by Stanek \etal (1998) turns out to be in excellent
agreement with our value.

In order to verify our results from the RC 
method, we have searched for a third, independent estimate of the
absolute $I_{0,m}$ of the RC in LMC field populations. For this aim
we have considered the data by Brocato \etal (1996) of a LMC region
around the old cluster NGC1786. The cluster reddening, as estimated by
Brocato \etal (1996) using the technique by Sarajedini (1994), results
to be E(B$-$V)=0.09$\pm$0.05, in agreement with the value derived by
Walker \& Mack (1988) for the field around the cluster.
We have then corrected the data
for extinction adopting the reddening law by Cardelli et
al.~(1989). The resulting $(V-I)_{0}$ color range spanned by the RC
is very much the same as in our data. We have then applied the
procedure described in section \S 2.1, obtaining
$I_{0,m}$=18.05$\pm$0.09~{\sl mag} (the contribution due to the reddening
uncertainty is included in the error), and $\sigma_{RC}$=0.17~{\sl mag}. The
results for both $I_{0,m}$ and $\sigma_{RC}$ are in good agreement with the
corresponding quantities we derived from our data. 

A remaining matter of concern appears to be the result by Udalski (1998) about the RC
level in 6 clusters of the LMC: SL388, SL663, SL862, NGC2121, NGC2155 and ESO121SC03).
These objects span the age range between 2 and 9 Gyr (suitable for comparisons with the 
field RC populations), and display an almost constant value 
of $I_{0,m}\simeq 17.9$~{\sl mag}. The extinctions are generally small, 
so that even an overestimate of the reddening cannot explain  
(at least not completely) the discrepancy. 
However, a deeper analysis of these clusters reveals that their brighter 
RC levels are in agreement with predictions from stellar evolutionary models 
and the ``long'' distance to the LMC.
More in detail, we have considered 5 of the mentioned clusters, for which the RC level is
determined with a reasonably large number of stars. We have excluded SL388 since 
the peak of its observed RC luminosity function is poorly populated and not sharply defined, 
but distributed over approximately 0.2 {\sl mag} (only 6 stars in the most populated bin 0.07 
{\sl mag} wide, 5 in the adjacent less luminous one, 5 again in the one 0.14 {\sl mag} brighter), 
thus making a statistical
determination of  $I_{0,m}$ not very reliable. According to Sarajedini (1998) SL663, NGC2121
and NGC2155 share the same metallicity (derived from the slope of the RGB, independently of
the assumed reddening), that is [Fe/H]$\simeq -1.0$. 
Bica \etal (1998) derived from Washington photometry of the RGB of
SL862 a metallicity [Fe/H]$=-0.9$, adopting a reddening E(B$-$V)=0.09 for this cluster.
By considering E(B$-$V)=0.12 as used by Udalski (1998) the derived metallicity lowers to 
[Fe/H]$=-1.0$ (Bica \etal 1998). Finally, in the case of ESO121SC03, Bica \etal (1998) find
[Fe/H]$=-1.05$ adopting E(B$-$V)=0.03, which becomes [Fe/H]$=-1.1$ if one
adopts the reddening used by Udalsky (1998), \ie E(B$-$V)=0.044.

In conclusion, all of the 5 clusters have approximately the same metallicity, 
[Fe/H]$\simeq-1.0$, which is at least a factor of 2 less than the average metallicity of the
field RC stars in our sample.
This fact helps in explaining the brighter RC levels found in the clusters. 
According to the models presented by Cole (1998) and Girardi (1999)
a metallicity difference $\Delta$[Fe/H]$\simeq$ 0.3 causes a difference of roughly
0.1 {\sl mag} in the RC level, the metal poorer one being brighter.
Taking into account this correction, possible small depth effects (these clusters are mainly
located in the halo of the galaxy) and the error budget, \ie the error associated to $I_{0,m}$
(typically a contribution by 0.02 mag due to the statistical error, and 0.03 mag 
of systematic error due to reddening uncertainties as estimated by Udalski 1998)
there is no serious contradiction between the LMC distance derived
from field stars or intermediate age clusters by means of the RC method.

In conclusion, we emphasize that our results based on different
and independent distance indicators seem to rule out the 
LMC distance evaluation recently provided by Stanek \etal (1998).
The recent revision by Zaritsky (1999) of the reddening for the  
fields analyzed by Stanek \etal (1998) further corroborates our result.
The present investigation represents an important evidence for the
paramount importance of carefully determining the reddening 
(and extinction) distribution for the stellar population one is planning to study.
We believe that additional work is needed in order to collect more reliable
estimations of both the mean value and the fluctuations of the interstellar extinction 
for the various stellar populations along the different lines-of-sight in the direction of the LMC.


\acknowledgements
We wish to warmly thank M. Lombardi for many precious suggestions, A.
Piersimoni for providing us with the data of the field around
NGC~1786, M. Groenewegen for many stimulating discussions about the
LMC distance, and Leo Girardi for fruitful discussions about the Red Clump
method. 
We wish to thank the anonymous referee for valuable comments that helped to
improve the presentation of the paper.
One of us (M.S.) would like to dedicate this paper to
the memory of the late Stanley Kubrick.


\pagebreak

\begin{figure*}
\centerline{
\psfig{figure=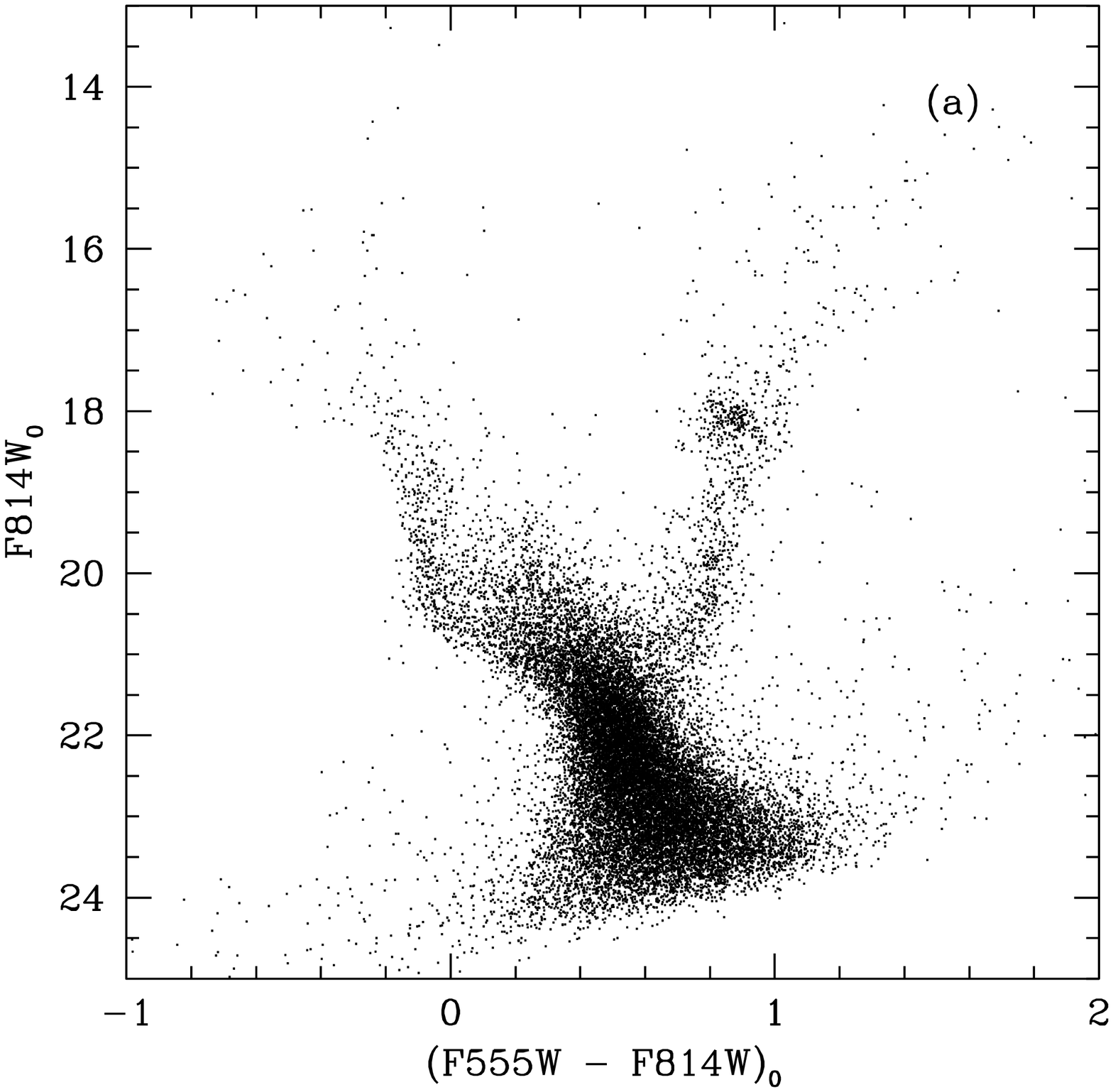,width=0.48\textwidth}
}
\label{fig1a}
\end{figure*}

\begin{figure*}
\centerline{
\psfig{figure=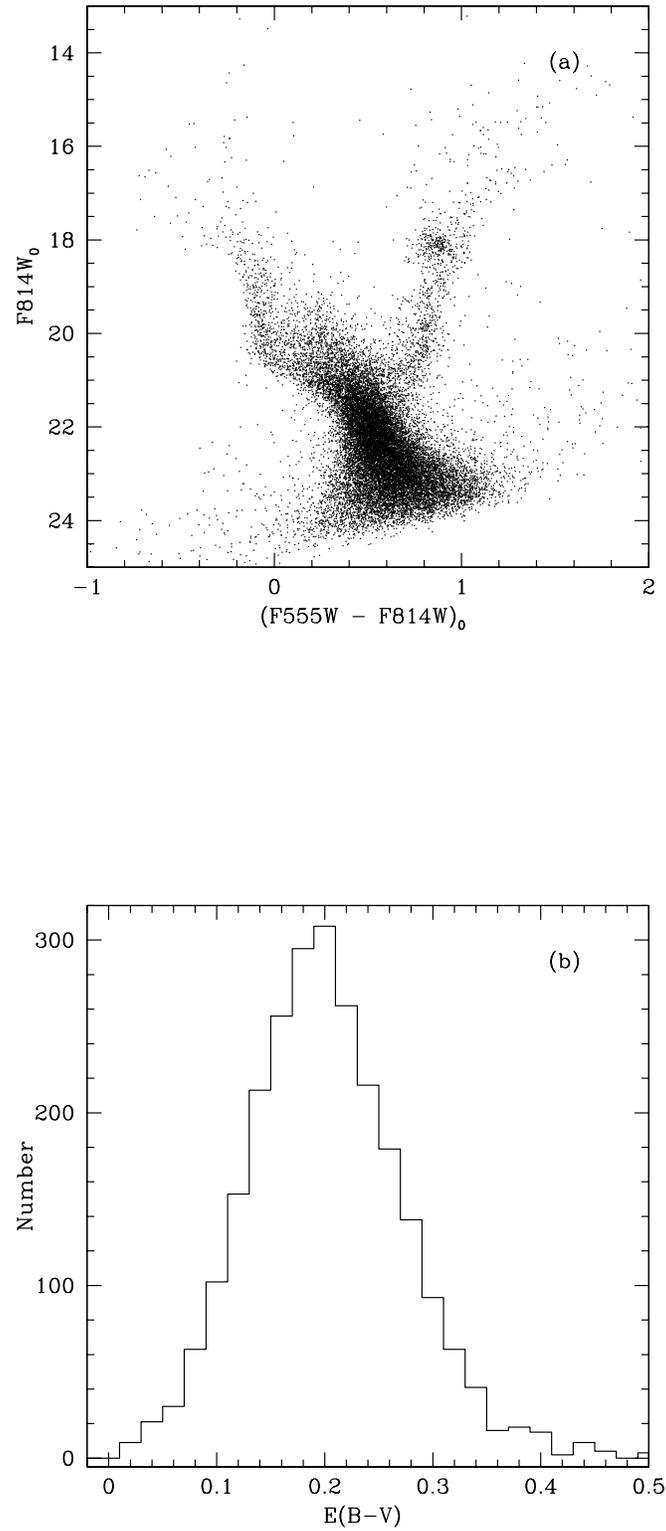,width=0.48\textwidth}
}
\caption{{\em Top panel}: Dereddened
F814W$_0$~vs.~(F555W$-$F814W)$_0$ Color-Magnitude Diagram of the field
around SN~1987A. 
{\em Bottom panel}: Reddening distribution in the
observed field.
}
\label{fig1}
\end{figure*}

\begin{figure*}
\centerline{
\psfig{figure=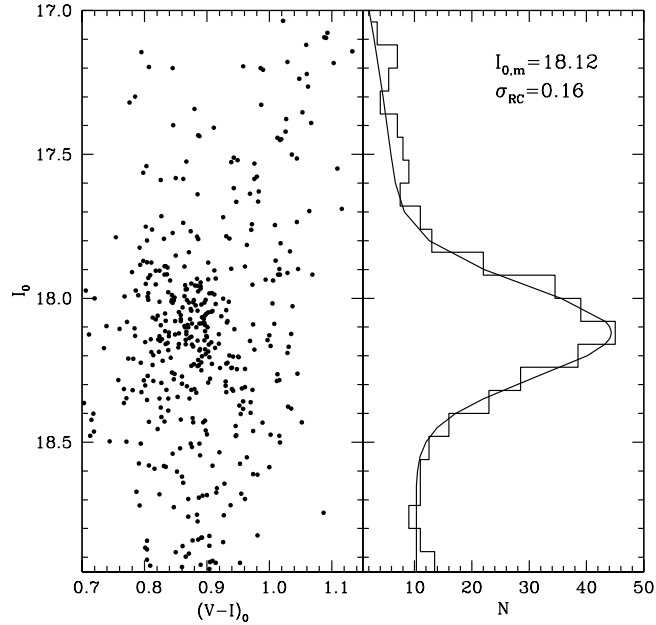,width=0.48\textwidth}
}
\caption{{\em Left panel}: The RC region in the CMD of the observed
LMC field. {\em Right panel}: Distribution of the RC stars as a
function of their $I_{0}$ magnitude, along with the analytical fit
as described by equation~1.
}
\label{fig2}
\end{figure*}

\begin{figure*}
\centerline{
\psfig{figure=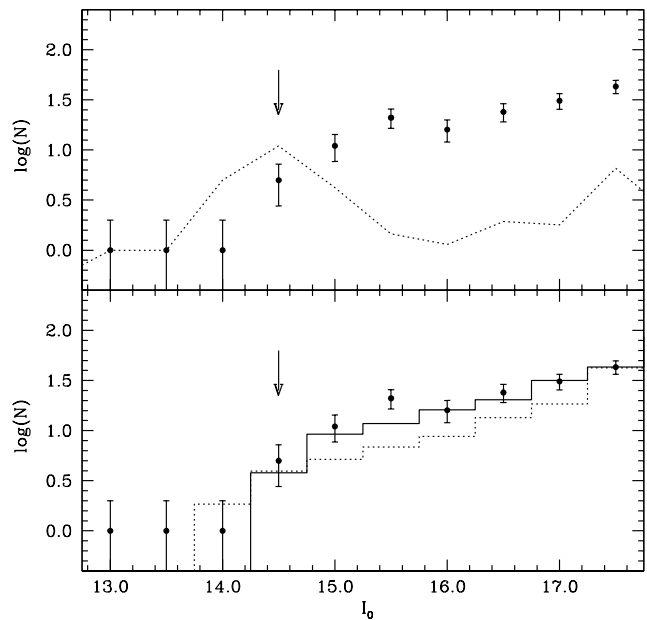,width=0.48\textwidth}
}
\caption{{\em Top panel}: The observational LF for the upper part
of the RGB (circles) and the convolution of the LF and the
edge-detector (dotted line). The arrow marks the position of the bin
where the TRGB discontinuity is detected. {\em Bottom panel}: Same as
the top panel but the theoretical LFs for 
$(m-M)_{0,TRGB}^{LMC}$=18.69 (solid line) and
$(m-M)_{0,TRGB}^{LMC}$=18.10 (dotted line) are also plotted.
}
\label{fig3}
\end{figure*}

\end{document}